\documentstyle[preprint,tighten,epsf,eqsecnum,aps,floats]{revtex}

\def\bea{\begin{eqnarray}}
\def\eea{\end{eqnarray}}
\def\endo{\end{document}}
\begin{document}
\draft
%\preprint{HEP/123-qed}

\preprint{%lc/talks/tabatinga.tex,
  NT@UW-00-29}
\title     {       Light Front Nuclear Theory and the HERMES Effect}
\author{Gerald A. Miller}
                                
\address{Department of Physics\\
  University of Washington\\
  Seattle, Washington 98195-1560}
%\lefthead{LEFT head}
%\righthead{RIGHT head}
\maketitle
\begin{abstract}
I discuss applications of relativistic
light front dynamics (the use of light cone variables)
to computing the nucleonic and mesonic
components of nuclear wave functions.  Our method is to
 use  a Lagrangian and its associated  energy-momentum
tensor $T^{^+\mu}$ to define the total momentum operators
$P^\mu$, with $P^+$ as the plus-momentum and $P^-$ the light cone time
development  %$\tau$-development
operator.
The aim  of this unusual approach to   nuclear physics is the desire
to use wave functions, expressed in terms  of plus-momentum variables,
 which are used to analyze high energy experiments  such
as deep inelastic scattering, Drell-Yan production, (e,e') and (p,p')
reactions.  We discuss or mention the topics:
infinite nuclear matter within the mean field
approximation; finite nuclei using the mean
field approximation; nucleon-nucleon scattering, within the one boson exchange
approximation; and, infinite nuclear matter including the effects of
two-nucleon correlations. Standard good results for nuclear
saturation properties are obtained, with a possible improvement in the lowered
 value, 180 MeV, of the  computed nuclear compressibility. In
 our approach,  manifest
 rotational invariance  can't be  used to simplify %as an aid in doing
 calculations. But for each
 of the examples reviewed here,
manifest rotational invariance  emerges at the end of the calculation.
Thus    nuclear physics can be done in
a manner in which modern nuclear dynamics is respected, boost
invariance in the $z$-direction is preserved, and in which 
rotational invariance
is maintained.  A salient feature is that $\omega,\sigma$
and $\pi$ mesons are obtained from the nuclear structure theory
as important constituents of nuclei.
I then  argue  that these constituents
can contribute coherently to enhance the electroproduction
cross section on nuclei for longitudinal virtual photons at low $Q^2$
while depleting the
cross section for transverse photons.   Thus the  recent HERMES
inelastic lepton-nucleus scattering data at low  $Q^2 $ and small
$x$ can be  described
using photon-meson and meson-nucleus couplings which are consistent
with (but not determined by) constraints  obtained from meson decay widths,
nuclear structure, deep inelastic scattering, and
lepton pair production data.    Our model makes a variety of
predictions testable in future experiments.
\end{abstract}
\section{Introduction}
This talk consists of two parts. The first is concerned with 
the subject of     light front nuclear theory\cite{gam}-\cite{miller00},
and the second with the 
application\cite{Miller:2000ta} of that theory to an effect
discovered in high energy lepton-nucleus scattering 
by the  HERMES collaboration\cite{Ackerstaff:1999ac}.
          
%end{document}
%\bigskip 
In this work, we will consider the regime in which the %the 
%\noindent 
nucleus  treated as being made of nucleons and mesons. Our goal is 
to       get the ground state wave function at zero temperature 
in terms of 
 light front coordinates. Using  these  coordinates to 
evaluate the consequences of a given Lagrangian is also called 
light front dynamics.
%\smallskip\noindent  goal--get wave function
%\bigskip\noindent%\smallskip %\noindent
We shall begin by giving a few more details concerned with answering the
following questions. 
What is light front nuclear theory?  Why do it?

We  shall present examples and results of  three 
studies:
%\smallskip\noindent
  infinite  nuclear matter treated using the  mean field approximation;
finite nuclei also treated with that approximation; and,
%\smallskip\noindent
  infinite   nuclear matter  including  correlations between two nucleons 
(light front Brueckner Theory).
In each application we find that 
%\bigskip\noindent
vector and scalar mesons are prominent  components      
 of  nuclear wave functions.

The second part  is concerned with % the idea that 
searching for  experimental consequences of these components.
Thus we discuss the                 
            HERMES effect\cite{Ackerstaff:1999ac}
            as a signature of these
nuclear mesons. This effect is observed in the interaction of      
            27.5 GeV positrons         with nuclei, and arises because
                     virtual photons have longitudinal $L$ polarization
as well as the usual transverse $T$ polarization.
The HERMES collaboration measures a ratio of cross sections for the
scattering of photons with different polarizations.  They find
\cite{Ackerstaff:1999ac}:
  \bea \left({\sigma_L(A)\over \sigma_L(D)} \right)/
\left(
    {\sigma_T(A)\over \sigma_T(D)}\right)
    \approx 5,\eea
for 
$x=0.01,$ and $ Q^2=0.5\; {\rm GeV}^2.$
This is truly a remarkable        result.  It has long been known that for
deep inelastic  scattering from a free nucleon one measures
$\left({\sigma_L\over \sigma_T}\right)\ll1.$ The vanishing of this
ratio is known (in other notation) as the 
Callan-Gross relation\cite{cagr},
and its verification provided the evidence for
the finding that the    partons observed in deep inelastic scattering
 are fermions (quarks). Now we have an experiment that seems to  
indicate that in nuclei the    partons are bosons.

%end{document}
\section  {     What is light front dynamics? }
                                
I try to mention only the most essential features. This dynamics is a
                                     relativistic many-body dynamics
in which 
%bigskip\bigskip\smallskip\noindent
the fields are quantized at a 
``time'', $t+z  =x^0+x^3\equiv x^+$     and the canonically conjugate 
 ``energy'' is given by $ p^0-p^3\equiv p^-$.
%\bigskip\bigskip\smallskip\noindent
Indeed, $p^-$ is the $x^+$ evolution operator, just as the Hamiltonian, $p^0$
is the  $x^0$ time evolution operator. 
%\bigskip\bigskip\smallskip\noindent
If $x^+$ is   ``time'', then for 
``space'' we must have
 $x^-\equiv t-z,$ with the canonically conjugate 
 ``momentum'' as   $p^+=p^0+p^3\equiv p^+.$ The other coordinates are treated 
as usual
$\vec{x}_\perp, \vec{p}_\perp.$ The use of the $p^+$ as the 
momentum variable is, for me, the reason behind the use of light front 
dynamics. This is because for a particle moving with large speed such that
 $\vec v\approx c\hat{e_3}, \;        p^+$     
is BIG, and for this reason is the experimentalists variable. Many observables 
are best expressed using this notation. In deep inelastic scattering the famous
Bjorken variable  
$x_{Bj}=Q^2/2M\nu$ is actually a ratio of plus momenta of the struck quark to
that of the entire struck nucleon.          
With our choice of variables: $ A^\pm\equiv A^0\pm A^3,$ and 
the dot product of two four vectors is given by 
\bea A\cdot B =A^\mu B_\mu={1\over2}\left(A^+B^- +A^-B^+\right)
-\vec{A}_\perp\cdot\vec{B}_\perp.\label{disp}\eea   
     The most important application of Eq.~(\ref{disp})  is the
 energy-momentum relation for a free particle:
\bea p^\mu p_\mu =m^2\quad=p^+p^--{ p_\perp}^2,\eea
so that \bea              
p^-={1\over p^+}
%\left
(p_\perp^2+m^2).\eea 
One has a          relativistic expression for the energy without
a   square root  operator. This is an enormous simplification 
when one wants to separate the coordinates of the  center-of-mass from the
rest of the wave function.
We may provide an approximate summary of light front dynamics:
Do ordinary quantum mechanics with energy denominators expressed 
in terms of $p^-$.

Another feature is that, when  one uses the Lagrangians of nuclear physics
the usual lore about light front dynamics should be true. That is the 
vacuum really is empty. This is because nucleons are heavy enough so that
nucleon pairs do not form vacuum condensates. Thus we should not ask what
the light front dynamics can do for nuclear physics. Instead we should 
ask what nuclear physics can do for light front dynamics. This is to provide
solutions of realistic, four-dimensional problems with relevance to 
observables.    
%\bigskip\bigskip \noindent Vacuum is empty

%Do ordinary quantum mechanics with energy denominators in terms of $p^-$.

%\end{document}
\section          {Motivation for Light Front Nuclear Physics}
Since much of this
work is
motivated by the desire to understand nuclear deep inelastic scattering
and related experiments,
it is worthwhile to review some of the features of the EMC
effect \cite{emc,emcrevs,fs2}. One key experimental result is the
suppression of the structure function for $x\sim 0.5$. This means that
the valence quarks of bound nucleons carry less plus-momentum than
those of free nucleons. Some other degrees of freedom must
carry the plus-momentum, and some authors therefore % This may be understood by
postulate that mesons carry a larger fraction of the plus-momentum in
the nucleus than in free space\cite{chls,et}.
While such a model explains the shift
in the valence distribution, one consequently obtains  a  meson (i.e.
anti-quark) distribution in the nucleus, which is  enhanced
compared to free nucleons, and which should be observable in Drell-Yan
experiments \cite{dyth}. However, no such enhancement has been observed
experimentally \cite{dyexp}, and  this has been termed as a severe crisis
for nuclear theory in
Ref.~\cite{missing}.

The EMC effect is rather small, so that one may begin by regarding the
nucleus as being made of nucleons. In this case, we say that 
deep inelastic scattering proceeds when a virtual photon is absorbed by
a quark carrying plus-momentum $p^+$, which came from a nucleon carrying
a plus-momentum $k^+$. In the parton model, the kinematic variable
$x_{Bj}=Q^2/2M_N\nu $ is given by 
\bea
x_{Bj}={p^+\over k^+}.\eea
Thus one needs to know
 the                 probability $f_N(k^+)$
that a nucleon has $k^+$. One also wants to know the related probability for a
meson, for example: $f_\pi(k^+)$.

Light front dynamics applies to nucleons
within the nucleus as well as to partons of the nucleons, and 
this is a useful approach whenever  the momentum of
initial or final state nucleons is large compared to their
mass \cite{fs}. For example,
this technique  can be used for $(e,e'p)$ and $(p,2p)$
reactions at sufficiently high energies. 

The essential technical advantage of using light cone variables
is that the light cone energy $P^-$ of a given final state 
does not appear in the delta function which expresses the
conservation of energy and momentum. Thus one may use closure to
perform the sum over final states which appears in the calculation of 
an exclusive nuclear cross section. The result is that the 
cross sections may be expressed in terms of the probabilities:
\bea
\sigma\propto f_N(k^+)\sim\int d^2k_\perp\cdots
\mid \Psi_{A,i}(k^+,k^\perp,\cdots)\mid^2,\eea where
$\Psi$ represents the ground state wave function.

For all these reasons we are concerned  with        calculating
 $f_N(k^+)$. One standard approach to the calculation, based on using the 
shell model equal time formulation is that: $E_\alpha +k^3\to k^+$. But
this can not be correct because $k^+$ is a continuous kinematic variable
which is not related to any discrete eigenvalues. Thus we need 
  realistic calculations, with real dynamics and
symmetries. This brings me to the conclusion 
 that it is necessary                  to
 redo nuclear physics on the light front.
%\end{document}

\section{    Light Front Quantization LITE}
%bigskip %\smallskip
Our attitude towards this topic is that
we need ${\cal L}$ no matter how bad!
This is because, in contrast with approaches based on symmetries,
we try to obtain all of the necessary operators from a given 
Lagrangian ${\cal L}$. The basic idea is to use the standard procedure to  
go from  ${\cal L}$ to $T^{\mu\nu}$, with the essential difference from
the usual procedure being that 
\bea P^\mu ={1\over 2}\int d^2x_\perp dx^- T^{+\mu}.\eea
A technical challenge arises because we need to express $ T^{+\mu}$
in terms of independent variables. For example, the  spin 1/2
 nucleon is always represented by 4 component spinor. Thus it has only
 2 independent degrees of freedom. One needs to express the two dependent 
variables in terms of the two
independent variables. This procedure is discussed in
the references.

We use two Lagrangians. The first is that of the  Walecka model\cite{bsjdw}:
  ${\cal L}(\phi,V^\mu,N)$ which contains the fields:
 nucleon $N$, neutral vector meson  $V^\mu$, neutral scalar meson $\phi$.
This is the simplest model which provides a reasonable caricature  of
the nucleus. The binding is caused by the attractive effects occurring at 
relatively long range 
when nucleons exchange scalar mesons. The nucleus  is prevented from collapsing
by the short distance repulsion arising from the exchange of vector mesons.

We also shall show results obtained using a more complicated chiral Lagrangian:
in which the fields are $  N,\pi,\sigma,\omega,\rho,\eta,\delta$. Our plan
is to  first use the 
     Walecka model in the       mean field approximation, and then 
to include $NN$
correlations using the chiral Lagrangian.
%\endo

\section { Infinite Nuclear Matter in  Mean Field Approximation-- Walecka
Model }

The basic idea behind the solution is very well known. One assumes that
the sources are strong, and produce   sufficiently many mesons to justify a
classical    treatment.          In infinite nuclear matter, one works in
a limit in which   the 
nuclear volume  $\Omega$ is considered to be  infinite, so that 
 all positions, and directions  equivalent in the nuclear rest frame.
In this limit the 
fields $\phi$ and $V^\pm$ become  constant, with    $V_\perp=0$. These features
simplify the solution of the 
           field equations. One
easily obtains the operators $
T^\pm$, and the light front ``momentum'' and ``energy'' are given by 
\bea {P^\pm\over\Omega}=\langle
T^\pm\rangle,\eea in which
 the expectation value is over  the nuclear ground state.

%\smallskip Self-consistent solution!

The  nuclear momentum content is the essential feature we
 wish to understand here. The results are that  \bea
{P^-\over\Omega}=m_s^2\phi^2+{4\over
(2\pi)^3}\int_F d^2k_\perp dk^+\;{k_\perp^2+(M+g_s\phi)^2\over k^+}\\
{P^+\over\Omega}=m_v^2(V^-)^2+{4\over
(2\pi)^3}\int_F d^2k_\perp dk^+\;k^+.\eea
The first term of $P^+$ is  the plus momentum carried by vector mesons,
and the second term is the plus momenta carried by the nucleons.
Here $g_s$ is the scalar-meson-nucleon 
coupling constant, and the vector meson-nucleon coupling constant 
$g_v$ enters in the expression for $V^-$.
The interpretation of these results is aided by a change of variables:
\bea
k^+\equiv \sqrt{(M+g_s\phi)^2+\vec{k}^2} +k^3,\eea  which
defines defines the variable $ k^3$. Using this variable one can show that 
rotational invariance is respected and obtain a spherical Fermi surface. 
Furthermore, one may show that  
$E\equiv{1\over 2}\left(P^-+P^+\right)$ is the           same   as
the usual expression obtained in the  Walecka  model.

  For nuclear matter in its rest frame we need to obtain
 $P^+=P^-=M_A$. This is the light front expression  of the statement that
the pressure on the system must vanish\cite{mm}. Indeed 
 the minimization  
\bea\left({\partial (E/A)\over\partial k_F}\right)_\Omega=0,\eea
determines the value of the Fermi momentum and is an expression
that gives %sets %\to\;k_F,\quad
$P^+=P^-=M_A$.  

We can quickly obtain the relevant numerical results using the 1974
parameters of 
 Chin \& Walecka. These are 
\bea g_v^2M_N^2/m_v^2=195.9\qquad g_s^2M_N^2/m_s^2=267.1.\eea
With these parameters 
$M_N+g_s\phi =0.56 M_N$ and $g_vV^-=270$ MeV. These are the  HUGE
scalar and vector potentials which are characteristic of the Walecka
 model.
The interesting variables are those associated with 
the  total nuclear plus momentum
 ${P^+\over \Omega}$. With  the above parameters, the 
vector meson contribution to this quantity: $%{P_V\over \Omega}=
m_v^2(V^-)^2$ is a monumental $0.35\;{P^+\over \Omega},$ while the nucleon
contribution  $
{4\over
(2\pi)^3}\int_F d^2k_\perp dk^+\;k^+$ is only 
$0.65{P^+\over\Omega}$. 
Only 65 \% of $P^+$ carried by nucleons,
but 90\% is needed to understand the EMC effect in infinite nuclear
matter\cite{sick}. 
This difference is huge. One can't plot the results of the theory in
comparison with the experimental results using pages of ordinary size.

So the large plus momentum carried by the vector  mesons is a feature of 
the Walecka model, which seems very odd. One needs to find means to
reduce this percentage from 35\%, but one also expects that within any model
 the 
vector mesons would carry some plus-momentum and it therefore becomes desirable
to search for an experimental verification of this feature. % signature of 

\section{ Beyond infinite  Nuclear Matter--  Mean Field Theory}

One possibility is that the large plus momentum carried by vector mesons
arises as  an artifact of using infinite nuclear matter.
We (Blunden,
Burkardt and I,\cite{fmf}) therefore investigated the subject
of light front Hartree Fock theory by performing 
 nucleus mean field theory calculations.

 The motivation is the need to
 find a way to reduce the plus-momentum carried by
 vector mesons, but one encounters a significant difficulty in the theory
 because the use of the variables $(x^-,\vec{x}_\perp)$ entails the loss
 of manifest rotational invariance. We found that the procedure of 
minimizing  $\langle P^-\rangle $ subject
to the constraint $\langle P^+\rangle= \langle P^-\rangle$, led to the recovery
of rotational invariance. This was seen by counting the number of
degenerate levels ($2j+1$)
that emerges from a numerical calculation in which
only rotational invariance about the $z$-axis could be used to simplify the
calculation. This is important for understanding the existence of
magic numbers in  nuclei.
The result of doing the  lengthy calculations was that 
nucleons were found (using the Walecka model)  to carry about 70\% of total
$P^+.$ This was only a modest improvement over the nuclear matter result of
65\%, and does not resolve the problem.

\section{  Beyond  Mean Field Theory}

The interactions between nucleons are strong, and the mean field approximation
is unlikely to provide a description of nuclear properties which involve
high momentum observables.  We developed \cite{gam,mm} a version of light
front theory in which the correlations between two nucleons are included.
The theory was applied to infinite  nuclear matter.

\begin{figure}
\unitlength1.1cm
\begin{picture} (21,11) (-1,2)
 \includegraphics{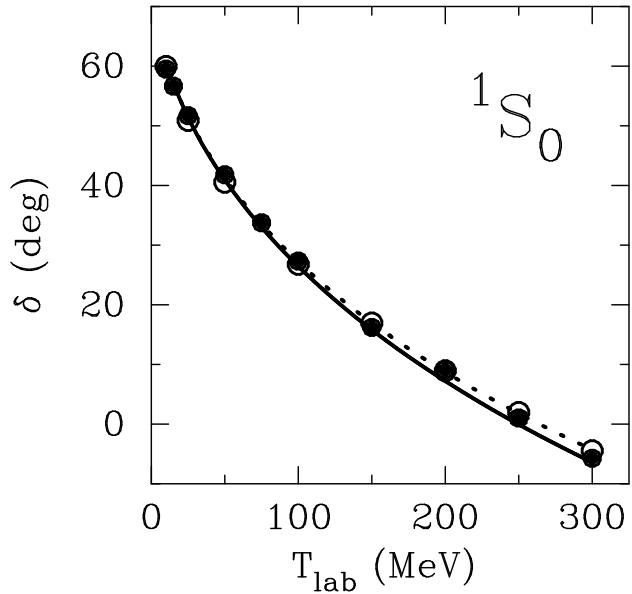}
\end{picture}
\caption{\baselineskip13pt Phase shift as a function of energy.
  Solid: our theory, % of Ref.~\cite{mm},
  dashed: an earlier relativistic
  theory.}
\label{fig:1s0}
\end{figure}

The calculation required three principal steps. (1)
 Light front quantization of
chiral ${\cal L}$. (2)
 Derive Light Front version of the $NN$ one boson
exchange potential. This could be done exploiting the relationship
between the Weinberg equation and the Blankenbecler-Sugar equation.
A sample of results for the phase shift is shown  in Fig.~\ref{fig:1s0}.
\begin{figure}
\unitlength.72cm %954cm
\begin{picture} (21,11)
  (1,2)
 \includegraphics{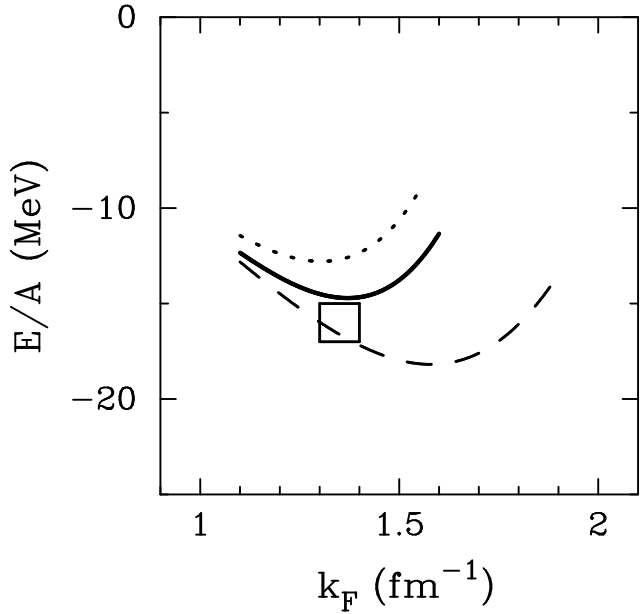}
\end{picture}
\caption{%\baselineskip13pt
  Nuclear matter binding energy per particle.
    Solid: our  theory, % of Ref.~\cite{mm},
  dashed: non-relativistic
  theory, dotted ignoring the effects of retardation.}
\label{fig:nucmat}
\end{figure}
The third step is to develop the many body theory. This turns out to be
a long story\cite{mm}. %, but the net result is that the formalism  looks
%just like ordinary non-relativistic many body formalism.
The net result is that the light front theory looks like the usual relativistic
Brueckner theory 
theory except that the Blankenbecler-Sugar equation is used, and the effects
of retardation are kept. The resulting nuclear matter saturation curve is
shown  in Fig.~\ref{fig:nucmat}.
 Standard good results for nuclear
saturation properties are obtained, with a possible improvement in the lowered
 value, 180 MeV, of the  computed nuclear compressibility.

 The results for deep inelastic scattering and the related Drell-Yan process
 seem very promising.  The nucleons carry at least 84\% of the nuclear plus
 momentum. This is calculated using only the Fermi gas part of the wave
 function. The percentage might increase to about 90\% (the target value for
 deep inelastic scattering) if the effects of the
 two-particle two-hole states are included. In these calculations the nucleus
 does have a pionic component, which arises as a result of going beyond teh
 mean field approximation. 
The number of excess pions per nucleon is about 5\%, and
although the distribution function of nuclear pions has not yet been computed,
this seems small enough to avoid a  contradiction with the Drell-Yan
data\cite{jm}. 

It seems that  nuclear physics can be done in
a manner in which modern nuclear dynamics is respected, boost
invariance in the $z$-direction is preserved, and in which the
rotational invariance so necessary to understanding the basic features
of nuclei is maintained.  A salient feature is that $\omega,\sigma$
and $\pi$ mesons are important constituents of nuclei. Another point, not
much dicussed here, is that it seems
possible to find
Lagrangians that yield reasonable descriptions of nuclear
deep inelastic scattering
and Drell-Yan reactions.

In the remainder of this talk I pursue the idea that the HERMES effect
provides a signature for the presence of nuclear $\omega$ and $\sigma$
mesons.

\section{Coherent Contributions of Nuclear Mesons to Electroproduction--
 HERMES
Effect }

%Gerald A. Miller,  S.J. Brodsky  M. Karliner 
%\smallskip \centerline{ hep-ph/0002156}
%\bigskip
%\centerline{\bf 

Let us discuss the nature of the   HERMES effect
\cite{Ackerstaff:1999ac}. Everyone is very familiar
with
the idea that lepton-nucleon or lepton nucleus scattering 
proceeds by the exchange of a virtual photon of four momentum
$q=(\nu,\vec{q})$. The important structure functions depend mainly
on $x_{Bj}$, with dependence on the logarithm of 
$Q^2=-q^2$. However, the cross sections depend on a third variable:
$y={\nu\over E}$, where $E$ 
is the energy of the incident lepton.
The    energy at which  HERMES runs, 27.5 GeV, is small enough so that
the experiment covers a wide range of values of $y$. This feature          
is what allows a new effect to be observed. In particular, the cross section
depends on the scattering of transversely $T$ and longitudinally $L$ polarized
photons\cite{roberts}:
\bea
\sigma\propto \sigma_T+\epsilon\sigma_L\\ %qquad
\epsilon\approx {4(1-y)\over 4 (1-y)+2y^2}.\eea
It is conventional to make the definition
\bea R\equiv
{\sigma_L\over \sigma_T}.\eea
In that case, one may manipulate the standard relations between the cross
section and structure function to find
\bea {\sigma_A\over \sigma_D}={F_2^A\over F_2^D}\quad
{1+\epsilon R_A\over1+R_A} \quad {1+R_D\over1+\epsilon R_D}.\eea
A linear dependence of the ratio on $\epsilon$ provides
a signature of a large value of $R_A$. Indeed, the 
  HERMES collaboration extracted  ${F_2^A\over F_2^D}$,and  $R_A$
from $x,Q^2,$ and $\epsilon$ dependence of ${\sigma_A\over \sigma_D}$.
The results are that 
\bea{\sigma_L(A)\over\sigma_L(D)}>1,\qquad
{\sigma_T(A)\over\sigma_T(D)}<1,\eea
with the largest effects 
\bea{R_A\over R_D}\approx 5\eea
obtained for $ x\approx 0.01,\quad Q^2=0.5\;{\rm GeV}^2.  $
As noted in the Introduction, this represents a huge violation of the  
 Callan-Gross relation\cite{cagr}, a violation large enough to indicate that,
 in nuclei,   bosons are the
partons of deep inelastic scattering! 
In the following, I describe the work of Ref.~\cite{Miller:2000ta}.

%\endo
\section{Nuclear Enhancement of $\sigma_L$}

We wish to describe the nuclear
enhancement of $\sigma_L$  and the nuclear suppression of $\sigma_T$
using a single input theory. We start with  $\sigma_L$. We found that
a  process in which a  virtual photon of four
momentum $q$
is converted in its interaction  with a nuclear $\omega$ meson into a scalar
meson of four momentum $p$, with $p^0\approx \nu$,
produces the desired enhancement.

To evaluate the effects of this process,
we need to determine the $\gamma\omega\sigma$ interaction.
We postulate a gauge-invariant form
\bea {\cal L}_I=
{g\;e\over 2 m_\omega}\;F^{\mu\nu}(\omega_\nu\partial_\mu \sigma
-\omega_\mu\partial_\nu \sigma) \label{lios} \eea
where $F^{\mu\nu}$ is the photon field strength tensor.
In momentum space one can use
\bea {\cal M}=
\displaystyle {g\over m_\omega}
\left(p\cdot q\; \epsilon^\gamma\cdot\epsilon^\omega-p\cdot
 \epsilon^\gamma \;q\cdot\epsilon^\omega\right)
F_V(Q^2)\label{mom}\eea
in which %$p_\mu$ is the momentum of the $\sigma$, and
we include a form factor $F_V$.  We seek
a constraint on the value of $g$ from the decay:
$\;\omega\to \sigma\gamma$.
The branching ratio for $\omega \to \pi^+ \pi^- \gamma <3.6\times 10^{-3}$
\cite{pdg},  which we assume to come from the process $\omega\to\gamma\sigma$
followed by the two pion decay of the $\sigma$ meson.
We may  determine an upper limit for g:
 $ {g_{UL}^2\alpha}=.013 \approx 2\alpha.\label{nom} $

\begin{figure}
\unitlength1.cm
\begin{picture}(15,9)(-3,-10)
 \includegraphics{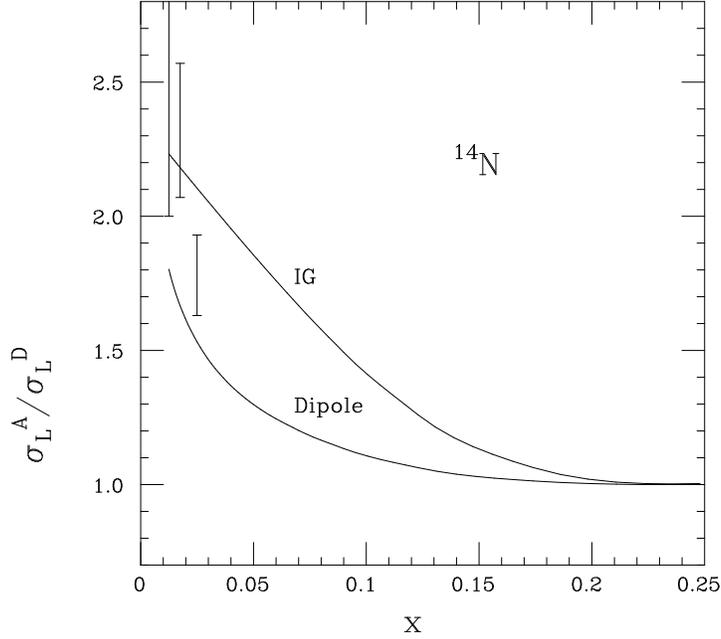}
\end{picture}
\caption{\baselineskip13pt ${\sigma_L(A)\over \sigma_L(D)}$, A=14, HERMES
data. 
The labels IG,  dipole refer to   form factors, see text.}
\label{fig:sla}
\end{figure}

Using this coupling constant, we obtained\cite{Miller:2000ta}
\bea\delta W^{00}\sim {(V^-)}^2 A^{1/3}\;\nu^3\;F_V^2(Q^2),\eea
in which $V^-$ is the value of the vector meson field at the center of the
nucleus. We use values which are
 consistent with nuclear saturation, and DIS, DY reactions.
 Notice the presence of the $\nu^3$ term which arises from the
factors of momentum appearing in Eq.~(\ref{mom}). This dependence is
essential because standard kinematics gives the result
\bea{\sigma_L(A)\over \sigma_L(D)}=1+{Q^4\over\nu( \nu^2+Q^2)}
{\delta W^{00}\over F_2^D R_D}
(1+R_D).\eea
The form factor $F_V(Q^2) $ is obtained from Ref.~\cite{ig}, we also use 
a dipole form factor. The results are shown in Fig.~(\ref{fig:sla}).
One is able to account for the large enhancement. 
%endo
\section{Nuclear Suppression of $\sigma_T(A)$}

Our explanation of the  nuclear  transverse cross section $\sigma_T$
data requires
a significant destructive interference effect at low $Q^2\approx 0.5-2\;
{\rm GeV}^2$. For small values of $x$, in the target rest frame, the
interaction proceeds by $\gamma^*$ decaying into a $q\bar q$ pair, which then
interacts with a target nucleon and emerges as a vector meson
$V_f$. For a nuclear target, we suppose that the
virtual photon interacts with a nuclear $\sigma$ meson and is converted
to an intermediate vector boson $V$, which is converted into the final
vector meson $V_f$ by a final state interaction.

We need to find the necessary $\gamma^*\sigma\to V$ interaction.
For consistency with
data taken at larger value of $Q^2$, we need an interaction
which decreases rapidly as $Q^2$ increases.  Furthermore,
the shadowing of the real photon ($Q^2=0$) is not very strong, and it is well
explained by conventional vector meson dominance models \cite{bauer}.
Thus consistency with all available data demands an amplitude for
$\gamma^*\sigma\to V$ which  vanishes, or is small,
as the $Q^2$ of the virtual photon $\gamma^*$ approaches 0.  This means that
 measuring the real photon decays of the vector mesons
provides no constraints on the coupling constant.

\begin{figure}
\unitlength1.cm
\begin{picture}(15,9)(-3,-10)
 \includegraphics{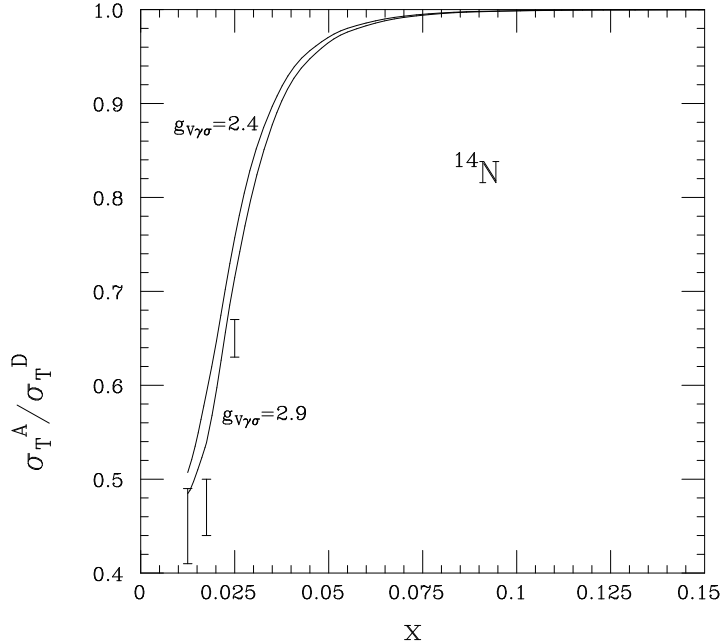}\hspace{1in}
\end{picture}
\caption{${\sigma_T(A)/ \sigma_T(D)}$, A=14, HERMES data      
                               }\label{fig:sta}
\end{figure}

We postulate the gauge-invariant interaction
\bea
 {\delta\cal L}_I=
\sum_V{g_{\gamma V\sigma}\;e\over 2 m_\sigma}\;F_{\mu\nu}\left[V^{\mu\nu}\sigma
+V^\mu\partial^\nu
\sigma
-V^\nu\partial_\mu \sigma\right]F_1^V,\eea
which in momentum space is  $\propto Q^2$. The details of the application
are given in the published work.
The results are shown in Fig.(\ref{fig:sta}).

\begin{figure}
\unitlength1.cm
\begin{picture}(15,9)(-3,-10)
 \includegraphics{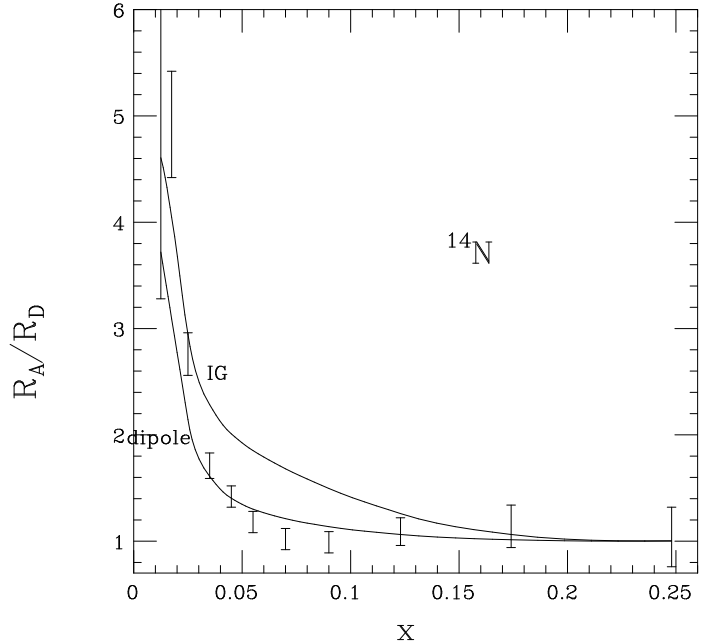}\hspace{1in}
\end{picture}
\caption{${R(A)/ R(D)}$, A=14 }
\label{fig:ra}
\end{figure}

 The nuclear enhancement of $R$ is obtained from
computing the
ratio of the previous results.  This is shown in Fig.(\ref{fig:ra}),
where it is seen one has a reasonably good description of the data.

We may summarize the results of our studies of the HERMES effect.
We find that $\sigma_L(A)$ is enhanced by the presence of  nuclear vector
mesons, and that $\sigma_T(A)$ is depleted by the presence of nuclear scalar
mesons. Both types of mesons are needed to obtain nuclei with 
correct binding energies  and densities. The values of the strong
coupling constants
used are also roughly consistent with data on nuclear deep inelastic scattering
data taken at larger values of $Q^2$.

Much verification of the present model is needed. 
Further tests of our model are possible.  An immediate consequence
would be the observation of  exclusive mesonic states
in the current fragmentation  region.  In particular, our description of
$\sigma_L(A)$ implies significant nuclear-coherent production of
$\sigma$ mesons along the virtual photon direction.  Our model for
the strong shadowing of coherent meson effects in $\sigma_T(A)$ can be tested
by measurements performed at the same value of $x$ but different values of
$Q^2$ than HERMES used.

The prospect that the mesonic fields which are responsible for nuclear binding
can be directly confirmed as effective fundamental constituents of nuclei at
small $x$ and  $Q^2\sim 1\;{\rm GeV}^2$
is an exciting development at the interface of traditional nuclear physics
and QCD.  The empirical confirmation of
nuclear-coherent meson contributions in the final state would
allow the identification of a specific dynamical mechanism for higher-twist
processes in electroproduction.
Clearly, these concepts should be explored further, both
experimentally and theoretically.

%\end{figure}^0$  is the vector potential for  the nucleon

\section*{Acknowledgments}

This work was supported in part by the United States
Department of Energy under contract 
and DE-FG03-97ER4104.  I thank all of the collaborators 
whose work I discuss here: P. Blunden, S.Brodsky, M. Burkardt, M.Karliner,
and R. Machleidt.

\end{document}